# A Cross-Whiskers Junction as a Novel Fabrication Process for Intrinsic Josephson Junction

Yoshihiko Takano [a,b,c], Takeshi Hatano [a,b,c,d], Akihiro Fukuyo [a,b,d], Akira Ishii [a,b], Shunichi Arisawa [a,b,c], Masashi Tachiki [a,b,c] and Kazumasa Togano [a,b,c]

[a] *National Institute for Materials Science, Sengen, Tsukuba 305-0047 Japan*
[b] *National Research Institute for Metals, Sengen, Tsukuba 305-0047 Japan*
[c] *CREST, Japan Science and Technology Corporation, Japan*
[d] *Fac. Sci&Tech, Science Univ. of Tokyo, Yamazaki, Noda 278-8510 Japan*

**Abstract**
A $Bi_2Sr_2CaCu_2O_{8+\delta}$ cross-whiskers junction has been successfully discovered as a novel intrinsic Josephson junction without using any technique for micro-fabrication. Two $Bi_2Sr_2CaCu_2O_{8+\delta}$ whisker crystals were placed crosswise on a MgO substrate and heated at 850°C for 30 min. They were electrically connected at their *c*-planes. The measurement terminals were made at the four ends of the whiskers. The I-V characteristics of the cross-whiskers junction at 5K were found to show a clear multiple-branch structure with a spacing of approximately 15 mV that is a feature of the intrinsic Josephson junction. The critical current density $J_C$ was estimated to be 1170 A/cm$^2$. The branch-structure was strongly suppressed by the magnetic field above 1kOe.

## Introduction

A high-temperature superconductor has a crystal structure in which $CuO_2$ layers are alternately stacked with blocking-layers. Josephson coupling between the $CuO_2$ layers was observed in $Bi_2Sr_2CaCu_2O_{8+\delta}$ high-$T_C$ superconductors in 1992 and called intrinsic Josephson junction because the $(BiO_2)$ layers work as insulating blocking layers intrinsically [1-3]. Furthermore, the theoretical prediction was proposed that the intrinsic Josephson device could generate terahertz-frequency electromagnetic radiation. Therefore, the intrinsic Josephson effect has received much attention ever since [4,5]. However, for the fabrication of intrinsic Josephson junctions, advanced micro-fabrication techniques, such as a focused-ion-beam (FIB) or high-resolution photolithography, are generally involved [6-8]. For the applications to the cryoelectronics and investigation into the mechanism of high-$T_C$ superconductivity [9], the development of a new and simple technique for the fabrication of intrinsic Josephson junctions is important [10]. We have successfully discovered a cross-whiskers junction made by joining two Bi2212 whiskers by post annealing. This is a new method for the fabrication of the intrinsic Josephson junction which does not require micro-fabrication. In this paper, we present the details of the fabrication process and the intrinsic Josephson properties of Bi2212 cross-whiskers junction.

Experimental

Among the high-$T_C$ superconductors, whisker crystals can be grown only in Bi-based system [1]. The general process of whisker preparation is in accordance with that of Matsubara *et al.* [11-12]. Whisker samples were prepared from 99.9% pure powders of $Bi_2O_3$, $SrCO_3$, $CaCO_3$, and $CuO$. These starting materials were mixed at a ratio of Bi:Sr:Ca:Cu=3:2:2:4. The obtained mixture was melted in an alumina crucible at 1200°C for 30min and subsequently quenched. The obtained glassy plate was placed in the alumina container and heated in an electric box furnace at 850°C for 120 hours with a flowing $N_2$-70%$O_2$ gas mixture. After the heat-treatment, many fibrous whisker crystals were grown from the surface of the glassy plate [13,14]. The obtained whiskers have a shape of narrow tape with *c*-axis perpendicular to the surface and *a*-axis parallel to the growth direction. The dimensions of the whiskers were as follows: width=10-30μm, thickness=1-3μm, length=$10^4$μm or more.

The fabrication process of a cross-whiskers junction is very simple. Figure 1 shows the schematic of the fabrication process. Two appropriate pieces of whiskers were placed crosswise on a MgO substrate. MgO(100) single-crystal was chosen as a substrate for the cross-whiskers junction because the thermal expansion of MgO is comparable to that of a Bi2212 superconductor. The *c*-planes of the whiskers were parallel to the substrate surface. Two whiskers were then attached to each other by their *c*-planes. To join the two whiskers, the obtained samples were annealed at 850°C for 30 min under a flowing gas mixture of $N_2$-70%$O_2$. For electronic transport measurements, gold wires were attached to the four ends of the cross-whiskers by silver paste. Because the surface of the whisker was insulating, the sample must be annealed again to obtain a low contact resistance. The resistance was measured by the four-probe method. Measurements of I-V characteristics were performed with a current-biased mode under various magnetic fields. A magnetic field was applied up to 1T along the *c*-axis. Figure 2 shows a photograph of the optical microscope of the cross-whiskers junction fabricated on a MgO substrate. In this picture, the crossed solid lines are whiskers, and the cross point is a junction. An scanning electron microscope image of the junction area is shown in Fig. 3. Whiskers play an important role as superconducting layers of the junction as well as the current leads.

Results and Discussion

Figure 4 shows the current-voltage (I-V) characteristics of the cross-whiskers junction measured at 5K without a magnetic field. The clear multiple-branch structure that is a feature of the intrinsic Josephson junction was observed. The magnitude of the first voltage jump is ~15 mV, which shows a good agreement with the typical one in the micro-fabricated intrinsic Josephson junction mesa structure [15]. The interval between the branches becomes smaller in the high-voltage region. A critical current at a zero-voltage branch is approximately 11 mA. From the critical current (11 mA) and junction area (938 μm$^2$), the critical current density $J_C$ at 5K was estimated to be ~1170 A/cm$^2$. This value is consistent with the typical $J_C$ observed in slightly over-doped Bi2212 intrinsic Josephson junctions [3]. We have observed clear intrinsic Josephson properties by joining two whiskers without micro-fabrications.

The I-V characteristics of the cross-whiskers junction were measured under various magnetic fields. Magnetic fields were applied parallel to the *c*-axis of the Bi2212 whiskers. Figure 5 displays the I-V characteristics measured under H=500 Oe. The branch structures were still clear, and the voltage interval in the branch structures were almost the same as those measured in the absence of a magnetic field. The magnitude of the $J_C$ was decreased by nearly 30%. On the other hand, a drastic change in branch-structure was observed above 1000 Oe. I-V characteristics measured at 5K under the field of 1000 Oe is shown in Fig. 6. The voltage jump and hysteresis were strongly suppressed by applying the field. The $J_C$ decreased by more than 50%. This field depndence of $J_C$ is comparable to the data observed in the single crystal [3]. The branch structure almost disappeared above H=2000 Oe.

The observation of a branch structure implies the existence of the regional mesa-like structure where the current flows along *c*-axis. It is not obvious how the mesa-like structure was formed between the two whiskers. We propose the following possibilities, model 1 and 2. Model 1: The surface of the as grown whisker used for cross-whiskers junction is insulating in nanometer thickness. Since the superconductivity will be recovered at cross-area by the heat treatment for joining two whiskers, inter whiskers superconducting current can pass through. The surfaces other than the cross area of the whiskers remain insulating. Then the cross area where the superconductivity recovered forms *c*-axis current path enclosed by the insulation layers, as shown Fig. 7. Thus, this superconducting cross-area work as mesa-like structure which suitable for intrinsic Josephson junction. Because this mesa-like structure connected top and bottom whiskers, these whiskers work as current lead to the mesa-like structure. Model 2: The recrystallization of superconductor may occur at the cross area of the whiskers by surface diffusion. A heating two crystals in close contact may results in a filling of the narrow gap between the whiskers by surface diffusion. The superconductivity may be recovered by the recrystallization, because this crossed area is no more at the surface of the whiskers. A mesa-like structure, as thick as the roughness of the whisker surface, will be formed between two whiskers in this model.

## 4. Conclusions

We have successfully fabricated a cross-whiskers junction and observed intrinsic Josephson effects by joining two Bi2212 whiskers with the use of an electric furnace. For the development of an application for superconductivity, such as a tera-herts generator, research for a simple fabrication process will be indispensable. The simple process developed in this study provides wide range of possibilities for the low temperature electronics researchers. Further more, a cross-whiskers junction could change its cross angle. Therefore, investigation of Josephson properties of this cross-whiskers junction will provide the fundamental keys for elucidating the mechanisms of high-Tc superconductivity.

## References
[1]  H. Maeda, Y. Tanaka, M. Fukutomi, and T. Asano 1988 Jpn. J. Appl. Phys. **27** L209.
[2]  R. Kleiner, F. Steinmeyer, G. Kunkel and P. Müller 1992 Phys. Rev. Lett. **68** 2394.
[3]  R. Kleiner and P. Müller 1994 Phys. Rev. B **49** 1327.
[4]  S. Koyama, and M. Tachiki 1995 Solid State commun. **96** 367.
[5]  M. Machida, T. Koyama, A. Tanaka and M. Tachiki 2000 Physica C **330** 85.
[6]  Yu. I. Latyshev and J. E. Nevelskaya 1994 Physica C **235-240** 2991.
[7]  S.-J. Kim, Yu. I. Latyshev and T. Yamashita 1998 Appl. Phys. Lett. **74** 1156.
[8]  A. Odagawa, M. Sakai, H. Adachi, K. Sentsune, T. Hirao and K. Yoshida 1997 Jpn. J. Appl. Phys. **36** L21.
[9]  Y. Takano, S. Takayanagi, S. Ogawa, T. Yamadaya and N. Mori 1997 Solid State Commun. **103** 215.
[10] Y. Takano, T. Hatano, A. Ishii, A. Fukuyo, Y. Sato, S. Arisawa and K. Togano, to be published in Physica C.
[11] I. Matsubara, H. Kageyama, H. Tanigawa, T. Ogura, H. Yamashita, and T. Kawai 1989 Jpn. J. Appl. Phys. **28** L1121.
[12] Y. I. Latyshev, I.G. Gorlova, A.M. Nikitina, V.U. Antokhina, S.G. Zybtsev, N.P. Kukhta and V.N. Timofeev 1993 Physica C **216** 471.
[13] T. Hatano, Y. Takano, A. Ishii, A. Fukuyo, S. Arisawa and K. Togano, to be published in Physica C.
[14] T. Hatano, Y. Takano, A. Fukuyo, S. Arisawa, A. Ishii and K. Togano, to be published in IEEE trans.
[15] M. Suzuki, T. Watanabe and A. Matsuda 1999 Phys. Rev. Lett. **82** 5361.

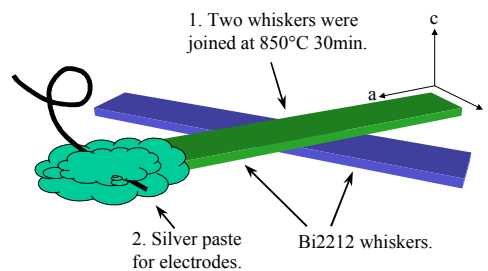

Fig.1 Schematic of fabrication process of the cross-whiskers junction.

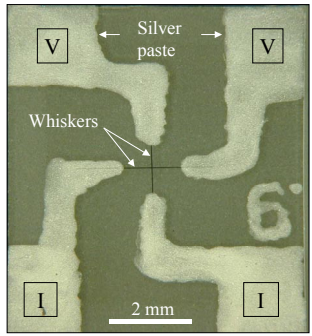

Fig.2 Photograph of the cross-whiskers junction fabricated on a MgO substrate.

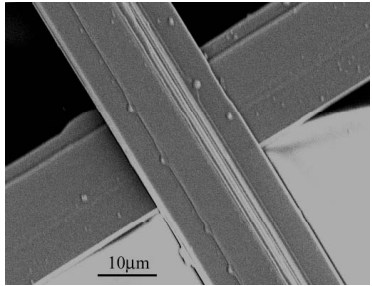

Fig.3 SEM image of the cross-whiskers junction.

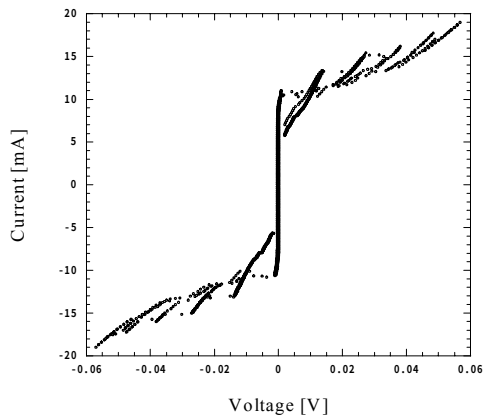

Fig.4 I-V characteristics of the cross-whiskers junction.

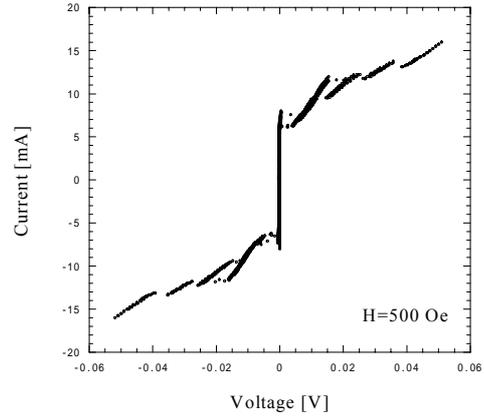

Fig.5 I-V characteristics of the cross-whiskers junction under H=500 Oe.

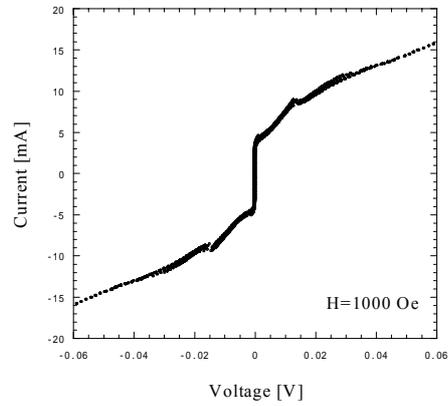

Fig.6 I-V characteristics of the cross-whiskers junction under H=1000 Oe.

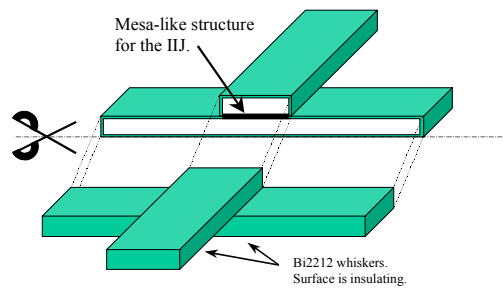

Fig.7 Schematic of the mesa-like structure formed around intersection of whiskers.